\documentclass[12pt]{JHEP3}
\usepackage{nicefrac}
\usepackage{graphicx}
\usepackage{amsmath}
\usepackage{amssymb}
\usepackage{latexsym}
\usepackage{subfigure}
\usepackage{epsfig}
\title{  Large $\bf{ N}$ Field Theory and AdS Tachyons}

\preprint{ YITP-SB-08-20}

\author{Elli Pomoni\footnote{Email: elli.pomoni@stonybrook.edu } $\,$  and Leonardo Rastelli\footnote{Email: leonardo.rastelli@stonybrook.edu}
\\ \\ \\
\it  C.N. Yang Institute for Theoretical Physics,\\
\it Stony Brook University, \\
\it Stony Brook, NY 11794-3840, USA}

\abstract{ 

\bigskip

In non-supersymmetric orbifolds of ${\cal N} =4$ super Yang-Mills, conformal invariance is broken by the {logarithmic} running of 
double-trace operators  -- a leading effect at large $N$. 
A tachyonic instability  in $AdS_5$
  has been proposed as the bulk dual of double-trace running.
  In this paper we make this correspondence more precise.
By standard field theory methods, we show that the double-trace 
beta function is  quadratic in the coupling,
to all orders in planar perturbation theory.
Tuning the double-trace coupling to its (complex) fixed point,
we find conformal dimensions of the form $2 \pm i \, b(\lambda)$,
as  formally expected for  operators dual to bulk scalars that violate the 
stability bound.
We also show that conformal invariance is broken in perturbation theory if and only
if dynamical symmetry breaking occurs.
Our analysis is applicable  to  a general large $N$ field theory with vanishing single-trace beta functions.
}

\newcommand{\Tr}{\mbox{Tr}}

\def\h{\eta}

\def\IC{\relax\hbox{$\inbar\kern-.3em{\rm C}$}}

\def\IC{{\bf C}}

\def\bea{\begin{eqnarray}}
\def\eea{\end{eqnarray}}
\def\be{\begin{equation}}
\def\ee{\end{equation}}
\def\ea{\end{align}}
\def\bse{\begin{subequations}}
\def\ese{\end{subequations}}
\def\1F1{{}^{(1)}\!F^{(1)}}
\def\2F0{{}_2\!F_0}

\def\sl{$SL(2,R)$}

\def\h3{$H_3^+$}

\def\IC{{\mathbb C}}

\def\lbldef#1#2{\expandafter\gdef\csname #1\endcsname {#2}}

\def\href#1#2{#2}

\newcommand{\beq}{\begin{equation}}
\newcommand{\eeq}{\end{equation}}
\newcommand{\ber}{\begin{eqnarray}}
\newcommand{\eer}{\end{eqnarray}}

\def\be{\begin{eqnarray}}
\def\ee{\end{eqnarray}}

\newcommand{\cO}{{\cal O}}
\newcommand{\bO}{\bar {\cal O}}

\def\sl{$SL(2,R)$}

\def\<{\langle}
\def\>{\rangle}


\begin{document}

\section{Introduction}

Conformal invariant quantum field theories in four  dimensions 
are interesting both theoretically and for potential phenomenological applications. 
While perturbatively finite supersymmetric QFTs have been known for a long time \cite{pertfinite}
and  a vast zoo of non-perturbative supersymmetric examples
 was discovered during the duality revolution of the 1990s,
only few non-supersymmetric, interacting CFTs in $d=4$
are  presently known.\footnote{Large $N$ Bank-Zaks \cite{Banks:1981nn} fixed points come to mind.}

\smallskip

The AdS/CFT correspondence \cite{Maldacena:1997re, Gubser:1998bc, Witten:1998qj} 
seems to offer an easy route to several more
examples. A well-known construction \cite{Kachru:1998ys, Lawrence:1998ja} starts by placing a stack of $N$ D3 branes
at an orbifold singularity  $\mathbb{R}^6/\Gamma$. In the decoupling limit one obtains
the duality between an orbifold of  ${\cal N}=4$ SYM 
by $\Gamma \subset { SU(4)}_R$ and Type IIB on $AdS_5 \times S^5/\Gamma$. Supersymmetry
is completely broken if $\Gamma  \not  \subset SU(3)$, 
but since the AdS factor of the geometry is unaffected by the orbifold procedure,
conformal invariance appears to be preserved, at least for large $N$.
However, in the absence of supersymmetry
one may worry about possible instabilities \cite{Bershadsky:1998mb}.

\smallskip

On the string theory side of the duality, one must draw a distinction \cite{Adams:2001jb}
according to whether the orbifold action has fixed points or acts freely on $S^5$.
 If $\Gamma$ has fixed points, there are always
closed string tachyons 
in the twisted sector. 
If  $\Gamma$ acts freely, the twisted strings are  stretched by a distance of the order of the $S^5$ radius $R$;
the would-be tachyons are then massive for large enough $R$
(strong  't Hooft coupling $\lambda$), but it is difficult to say anything definite about small $R$. 

\smallskip

On the field theory side, a perturbative analysis
 at small $\lambda$ reveals that  conformal invariance is {\it always} broken, regardless
 of whether the orbifold is freely acting or not \cite{Dymarsky:2005uh, Dymarsky:2005nc}.
The inheritance arguments of \cite{Bershadsky:1998mb, Bershadsky:1998cb} guarantee that
the orbifold theory is conformal in its single-trace sector: at large $N$,
all couplings of marginal single-trace operators
have vanishing beta functions. 
However, even at leading order in $N$, there are
 non-zero beta functions for double-trace couplings of the form
\be \label{deltaS}
\delta S =  f  \int d^4 x \,  {\cal O}  \bar {\cal O}\, , \label{form}
\ee
where ${\cal O}$ is a twisted single-trace operator of classical dimension two \cite{Tseytlin:1999ii, Csaki:1999uy, Adams:2001jb, Dymarsky:2005uh, Dymarsky:2005nc}.
Conformal invariance could still be restored, if all  double-trace couplings $f_k$
had conformal fixed points.  It turns out that  this is never the case in the one-loop approximation \cite{Dymarsky:2005uh, Dymarsky:2005nc}. 
So for sufficiently small 
$\lambda$, all non-supersymmetric orbifolds of ${\cal N}=4$ break
conformal invariance.

\smallskip

\begin{figure}[t]
   \hspace*{20cm}
  \begin{center}
   \epsfxsize=8cm
   \epsfbox{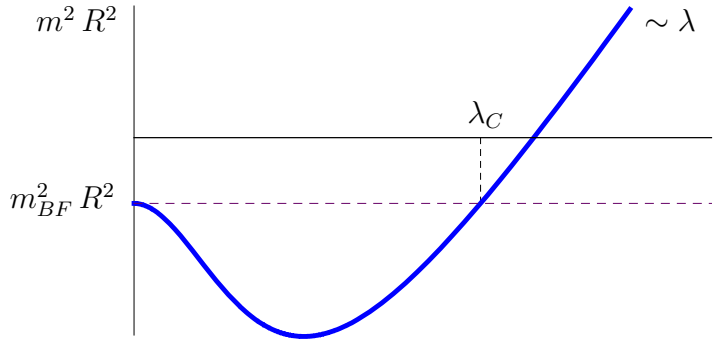}   
         \put(-260,120){$m^2\,R^2$}     
      \put(-270,53){$m_{BF}^2\,R^2 $}         
      \put(-30,120){$\sim \lambda$}
              \put(-97,85){$ \lambda_{C}$}
   \end{center} 
 \caption[x]{ \it Proposal for the qualitative behavior of  a ``tachyon'' mass in a freely acting
 orbifold, as a function of the 't Hooft coupling $\lambda$. The field is an actual tachyon (violating
 the BF stability bound) for $\lambda < \lambda_C$. See section \ref{examples} for more comments.}
   \label{proposal}
  \end{figure}

It is natural to associate this breaking
of conformal invariance with the presence of tachyons in the dual AdS theory \cite{Dymarsky:2005uh}. By an AdS tachyon, we mean a scalar field that {\it violates} the 
Breitenlohner-Freedman bound~\cite{Breitenlohner:1982bm}:
\be
 {\rm For \;  a \; tachyon}\,,  \quad m^2 < m^2_{BF} = - \frac{4\,}{R^2}  \,.\qquad 
\ee
One is then led to speculate \cite{Dymarsky:2005uh} that
even for freely acting orbifolds, some of the twisted states must become tachyonic
for $\lambda $ smaller than some critical value $\lambda_C$. The conjectural behavior
of $m^2 (\lambda)$ for a  ``tachyon''  in a freely acting orbifold theory is shown in Figure \ref{proposal}.
A related viewpoint \cite{Adams:2001jb} links the tachyonic instability in the bulk theory with a perturbative Coleman-Weinberg  instability in the boundary
theory. From this latter viewpoint however, it seems at first that whether $\Gamma$ is freely acting or not
makes a difference even at weak coupling \cite{Adams:2001jb}: if $\Gamma$ has fixed points, the quantum-generated double-trace potential
  destabilizes the theory along a classical flat direction;
 if $\Gamma$ is freely acting, the symmetric vacuum appears  to be stable, because  twisted operators
 have zero vevs along classical flat directions.
 
\smallskip

In this paper we make the correspondence between double-trace running and bulk tachyons more precise.
Taken at face value, an $AdS_5$ tachyon would appear to be dual to a boundary operator with {\it complex} conformal dimension
of the form
\be
\Delta = 2 \pm i \, b \, , \quad b = \sqrt{ |m^2 R^2 + 4|} \, .
\ee
We are going to find a formal sense in which this is correct, and
a prescription to compute the tachyon mass $m^2 (\lambda)$ from the  boundary theory.  In principle this 
prescription   could be implemented order by order in $\lambda$ and allow to test  the conjectural picture of Figure \ref{proposal}.
We also show that  the perturbative
 CW instability is  present if and only if conformal
invariance is broken, independently of the tree-level potential, and thus independently of whether the
orbifold is freely acting or not.

\smallskip

Our analysis applies to the rather general class of large $N$
theories ``conformal in their single-trace sector''. We consider non-supersymmetric,
classically conformal field theories with  lagrangian of the standard  single-trace form ${\cal L} = N \, {\rm Tr} \, [ \dots \, ] $.
Denoting collectively by $\lambda$  
the single-trace couplings that are kept fixed in the large $N$ limit,\footnote{
In the example of an orbifold of ${\cal N}=4$ SYM, $\lambda  = g^2_{YM} N$ 
is the usual 't Hooft coupling.} we assume that $\beta_\lambda \equiv \mu \frac{\partial}{\partial \mu }\lambda = 0$ 
at large $N$.  Generically however, perturbative renormalizability
forces  the addition of double-trace couplings of the form (\ref{form}), where
${\cal O} \sim {\rm Tr}\,  \phi^2$ is a single trace operator of classical dimension two. 
Thus it is essential to compute the double-trace beta functions $\beta_f$  to
determine whether or not conformal invariance is maintained in the quantum theory.
Our main technical results are expressions for  $\beta_f$,
 for the conformal dimension $\Delta_{\cal O}$ and for the effective potential ${\cal V}(\varphi)$,
valid to all orders in  planar perturbation theory.

\smallskip

Besides orbifolds of ${\cal N} = 4$ SYM,  other examples of large $N$ theories conformal in their single-trace sector
 are certain non-supersymmetric continuous deformations 
of ${\cal N}=4$ SYM \cite{Lunin:2005jy, Frolov:2005dj, Ananth:2007px}. One can also contemplate theories with adjoint {\it and} fundamental
matter, where the instability arises in the mesonic sector and is dual to an {\it open}
string tachyon. A detailed analysis of such an ``open string'' example will appear in a forthcoming paper \cite{paper2}.
Somewhat surprisingly, conformal invariance turns out to be broken
in all concrete cases of  non-supersymmetric ``single-trace conformal''  theories  that have been studied so far. There is no a priori reason
of why this should be the case in general. A  more systematic search for conformal examples is certainly warranted.

\smallskip

We should also mention from the outset that independently of the perturbative instabilities which are the focus of this paper,
 non-supersymmetric orbifold theories may exhibit  
a non-perturbative instability akin to the decay of the Kaluza-Klein vacuum \cite{Horowitz:2007pr} (see also \cite{Copsey:2008fs}).
For a class of  freely acting $\mathbb{Z}_{2 k +1}$ orbifolds, at large coupling $\lambda$
the decay-rate per unit volume scales as \cite{Horowitz:2007pr}  
\be
\Gamma_{decay} \sim k^9 e^{-N^2 /k^8} \Lambda^4 \, ,
\ee
where $\Lambda$ is a UV cut-off.  This instability is logically distinct and parametrically
different from the tree-level tachyonic instability. 
 It is conceivable that a given orbifold theory
may be stable in a window of couplings $\lambda_C < \lambda < \lambda_{KK}$
intermediate between a critical value $\lambda_C$ where the
``tachyon'' is lifted (Figure \ref{proposal}) and another  critical value $\lambda_{KK}$ where the the non-perturbative instability {sets~in}.

\smallskip

Multitrace deformations in the context of the AdS/CFT correspondence have been investigated in several papers,
beginning with \cite{Aharony:2001pa, Aharony:2001dp, Witten:2001ua,  Berkooz:2002ug, Minces:2002wp}.

\smallskip

The paper is organized as follows.
In section \ref{betagamma} we study the renormalization
of  a general field theory conformal in the single-trace sector 
and derive expressions for  $\beta_f$ and ~$\Delta_{{\cal O}}$ valid to all orders in planar perturbation theory.
In section \ref{CW} we study the behavior of the running coupling $f(\mu)$
and the issue of stability of the quantum effective potential ${\cal V}(\varphi)$. 
In section \ref{prescription} we make our proposal for the computation of the  tachyon mass $m^2(\lambda)$
from the dual field theory. We illustrate the prescription in a couple of examples 
and make some remarks on flat directions in freely acting orbifold theories.
We conclude in section \ref{discussion} discussing a few  open problems.

\section{Renormalization of double-trace couplings}

\label{betagamma}

We are interested  in large $N$, non-supersymmetric field theories in four dimensions.
We start  with a conformally invariant classical action of  the standard single-trace  form.
Schematically,
\be \label{Sst}
S_{ST} [N, \lambda] =  \int d^4 x \, N \, {\rm Tr} \, [   (D \phi)^2 +  \psi  D \psi  + (DA)^2 + \lambda \, \phi^4 + \dots ] \, ,
\ee
where $\phi$, $\psi$, $A$  are $N \times N$ matrix-valued scalar, spinor and gauge fields. We have
written out the sample interaction term $N \lambda  \Tr \, \phi^4$ to establish our notation for the couplings: we denote 
collectively by $\lambda$ the couplings in $S_{ST}$ that are kept fixed in the large $N$ limit.
 
 \smallskip

 \begin{figure}[h] 
  \centering 
\mbox{\subfigure[]{\epsfig{figure=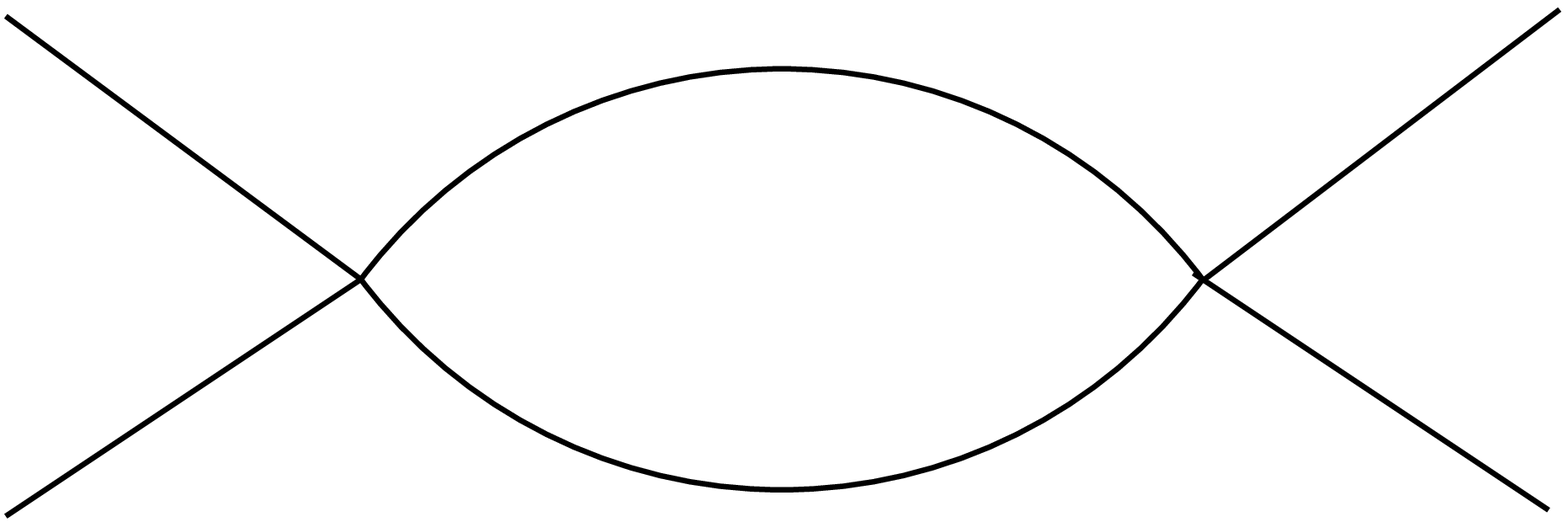,width=5cm,angle=90}}\quad  \quad \quad  \quad  \quad \quad
\subfigure[]{\epsfig{figure=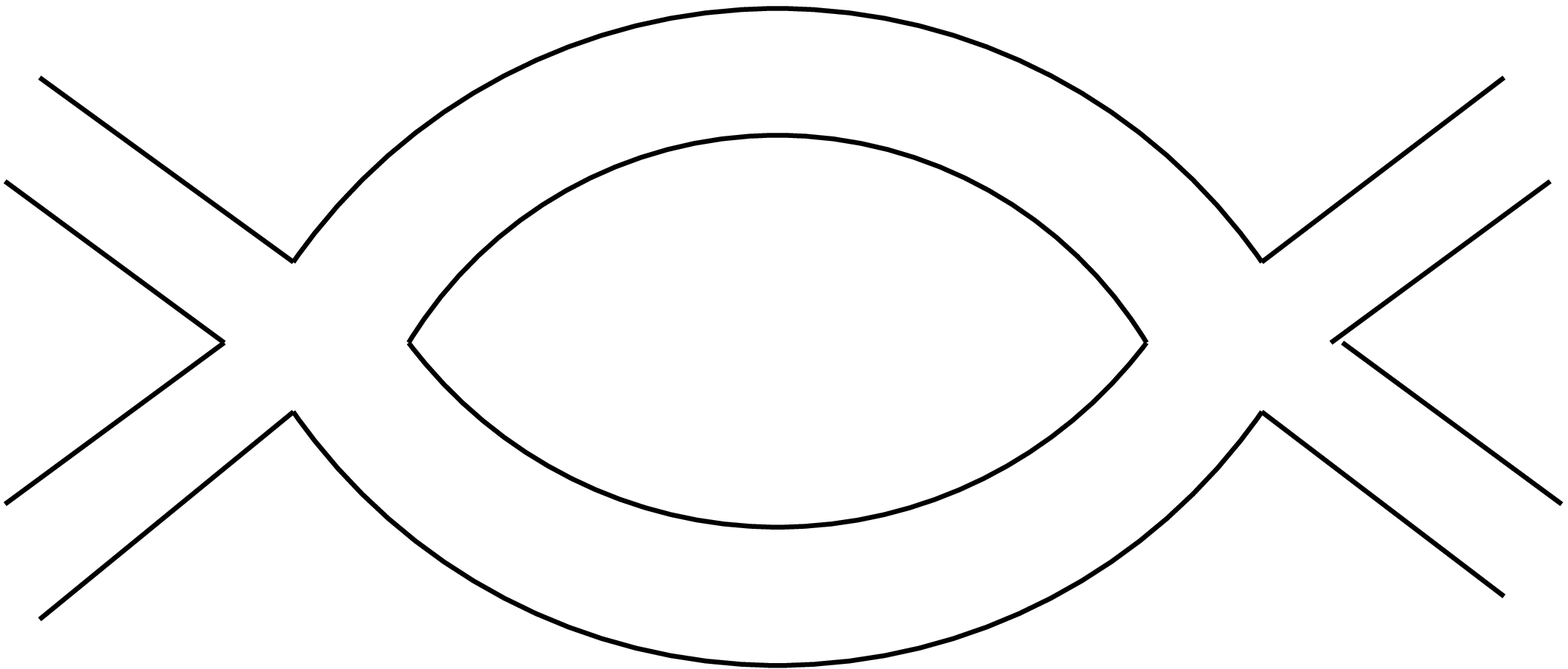,width=5cm,angle=90}}\quad  \quad  \quad  \quad  \quad \quad
      \subfigure[ ]{\epsfig{figure=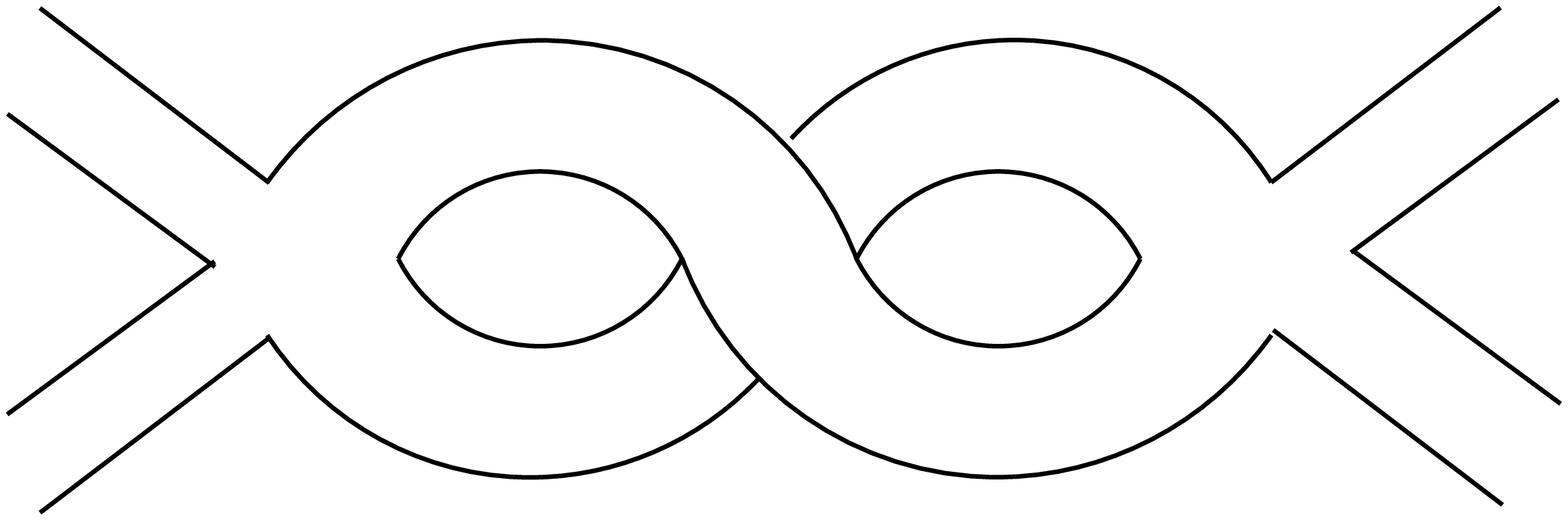,width=5cm,angle=90}}} 
        \put(-260,110){$\lambda N$}
  \put(-260,30){$\lambda N$}
    \put(-250,70){$\frac{1}{ N}$}
      \put(-300,70){$\frac{1}{ N}$}
        \put(-138,-17){$  N \,\Tr \phi^4  $}
          \put(-15,-17){$(\Tr \phi^2 )^2$}
 \caption{  \it{ One-loop contributions to the effective action from a diagram with two quartic vertices.  Each
 vertex contributes a factor of $\lambda N$ and each propagator a factor of $1/N$, as indicated in (a). There
  are two ways to contract color indices: a single-trace structure 
 (b), or a double-trace structure~(c). }   } 
  \label{effective} 
\end{figure}
  Generically, the action (\ref{Sst}) is not renormalizable as it stands, because extra double-trace interactions are induced
 by quantum corrections. It is an elementary but under-appreciated fact that double-trace renormalization
is a {\it leading} effect at large $N$. For example,  consider  the contribution to the effective action from
one-loop diagrams  with two quartic scalar vertices~(Figure~\ref{effective}). Schematically,
\be \label{counting}
 \int  d^4 x \,  N\, \lambda   \Tr \,  \phi^4(x) \; \int d^4 y \,  N \, \lambda  \Tr  \,  \phi^4(y)    \,
 \sim   \lambda^2 \,   \log \Lambda \, \int d^4 z \,  \left[ \,   N \, \rm Tr \, \phi^4 +  (\rm \, Tr \, \phi^2   )^2 \,  \right] \,.
\ee   \label{}
The single-trace term $N \, \rm Tr \phi^4$ renormalizes a coupling already present
in the action (\ref{Sst}). The double-trace term $ (\rm Tr \phi^2 )^2$ 
 forces the addition of an extra piece to the bare action,
\be
S = S_{ST} + S_{DT} \, , \qquad S_{DT} = \int d^4x \, f_0 \,  ( \Tr \phi^2 )^2  \,,  \qquad f_0 \sim \lambda^2 \log \Lambda \,.
\ee
It is crucial to realize that $S_{ST}$ and $S_{DT}$ are of the same order 
in the large $N$ limit, namely $O(N^2)$. For $S_{ST}$, one factor of $N$ is explicit and the other arises from
the trace; for $S_{DT}$, each trace contributes one factor of $N$.

\smallskip

In the following, we specialize to theories for which the single-trace couplings do not run in the large
$N$ limit, $\beta_\lambda = \mu \frac{\partial}{\partial \mu} \lambda = O(1/N)$. In particular 
 the single-trace contribution in (\ref{counting}) is canceled when we add all the relevant Feynman diagrams. 
 This is what happens  in orbifolds of ${\cal N}=4$ SYM.
Twisted single-trace couplings cannot be generated in the effective action, since they are charged under the quantum symmetry,
while  untwisted single-trace couplings are not renormalized, since
they behave as in the parent theory by large $N$ inheritance.
However, neither argument applies to double-trace couplings of the form
$ f \, {\cal O}_g \,  {\cal O}_g^\dagger \,$, where  ${\cal O}_g =  {\rm Tr} (g  \phi^2)$ is
a twisted single-trace operator of classical dimension two.\footnote{ Here $\Tr =  \Tr_{SU(|\Gamma| N) }  $  and
$g \in \Gamma$. } Such double-trace couplings will be generated in perturbation theory.

\smallskip
In this rest of this section, we analyze the general structure of  double-trace renormalization.

\subsection{Double-trace renormalization to all orders}

The beta function for the double-trace coupling (\ref{deltaS}) was computed at one loop in \cite{Dymarsky:2005uh}, 
\be \label{betaf1}
\beta_f \equiv \mu \frac{\partial}{\partial \mu} f = v^{(1)} f^2 + 2 \gamma^{(1)} \lambda f + a^{(1)} \lambda^2 \,.
\ee

\begin{figure}[t]
 \centering
\mbox{\subfigure[]{\epsfig{figure=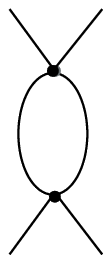,width=1.5cm,angle=0}}\quad
\quad  \quad  \quad \quad  \quad  \quad
     \subfigure[]{\epsfig{figure=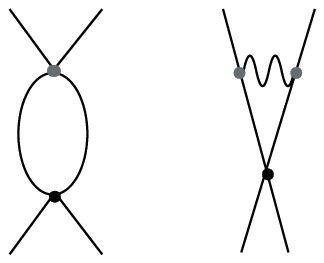,width=4.4cm,angle=0}}
    \quad  \quad\quad \quad  \quad  \quad
     \subfigure[]{\epsfig{figure=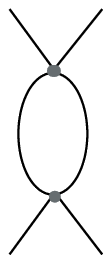,width=1.5cm,angle=0}}}
   \put(3,80){$\lambda$}
       \put(3,20){$\lambda$}
   \put(-118,75){$\lambda^{1/2}$}
\put(-174,75){$\lambda^{1/2}$}
    \put(-126,27){$f$}
 \put(-248,80){$\lambda$}
   \put(-248,20){$f$}
    \put(-378,80){$f$}
   \put(-378,20){$f$}
\caption{\it Sample diagrams contributing to $\beta_f$ at one loop: (a) $v^{(1)}f^2 \;$; (b) $2 \gamma^{(1)}\lambda f \; $; (c) $a^{(1)}\lambda^2$.}
 \label{oneloop}
\end{figure}

This result applies  to any theory conformal in its single-trace sector.
Here $v^{(1)}$ is the normalization of the single-trace operator ${\cal O} \sim \Tr \, \phi^2$, defined as
\be
\langle {\cal O}(x)  \bar {\cal O}(y)  \rangle = \frac{v^{(1)}}{ 2 \pi^2 (x-y)^4} \, .
\ee
The quantity $\gamma^{(1)} \lambda$ is the one-loop contribution to
the anomalous dimension of ${\cal O}$ from the single-trace interactions. 
The double-trace interaction also contributes to the renormalization of ${\cal O}$,
so that the full result for its one-loop anomalous dimension
 is
\be
\gamma_{\cal O}= \gamma^{(1)} \lambda + v^{(1)} f \, .
\ee
Some representative Feynman diagrams contributing to $\beta_f$  are shown in Figure \ref{oneloop}.
Our goal is to generalize these results to all orders in planar perturbation theory.

\subsubsection{The $\lambda = 0$ case}

\begin{figure}[t]
   \hspace*{10cm}
  \begin{center}
   \epsfxsize=8cm
   \epsfbox{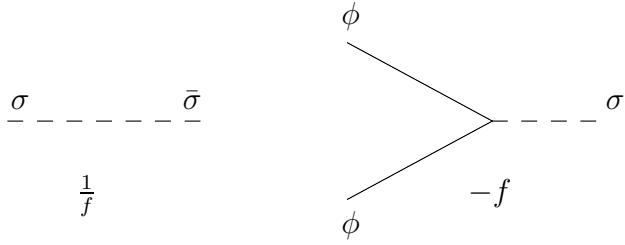}   
   \put(-225,35){$\sigma$}     
      \put(-160,35){$\bar{\sigma}$}   
  \put(0,35){$\sigma$}     
   \put(-100,67){$\phi$}  
      \put(-100,-12){$\phi$}   
       \put(-200,0){$\frac{1}{f}$}    
              \put(-52,0){$-f$}   
   \end{center}
 \caption[x]{ \it Feynman rules for (\ref{Ltree}).}
   \label{feynman}
  \end{figure}

\begin{figure}[t] 
  \centering 
\mbox{\subfigure[]{\epsfig{figure=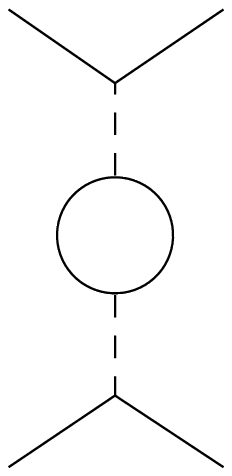,width=1.8cm,angle=0}}\quad  \quad \quad  \quad  \quad \quad
\subfigure[]{\epsfig{figure=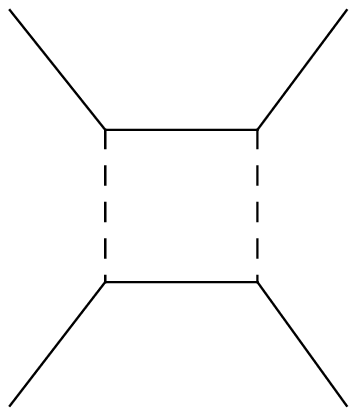,width=3cm,angle=0}}} 
 \caption{\it Diagram (a) is leading at large $N$, of order $O(1)$, but it is reducible. Diagram (b) is irreducible but it is subleading at large $N$, of order $O(1/N^2)$. } 
  \label{subleading} 
\end{figure}

Let us first practice with the simple situation where the single-trace part of the action is free.\footnote{ The calculation of $\beta_f$ for this case
already appears in \cite{Witten:2001ua}.} The total lagrangian is
\be
{\cal L}= {\cal L}^{free}_{ST} + {\cal L}_{DT} \, , \qquad {\cal L}_{DT} = f    \, {\cal O}   \bar {\cal O} \,  .
\ee
The discussion of the large $N$ theory is facilitated by a Hubbard-Stratonovich transformation.
We introduce the auxiliary complex scalar field $\sigma$ 
and write the equivalent form for the double-trace interaction,\footnote{For ease of notation we suppress possible flavor indices for ${\cal O}$ and $\sigma$.}
\be \label{Ltree}
{\cal L}_{DT}  = -f \sigma \bar \sigma + f \sigma \bar {\cal O} +f  \bar \sigma {\cal O} \,.
\ee
The obvious Feynman rules are displayed in Figure \ref{feynman}. 
The renormalization program is carried out as usual, by adding  to the tree-level lagrangian (\ref{Ltree}) local counterterms,
which we parametrize as
\be 
\delta {\cal L}_{DT} = - (Z_2 -1) f \sigma \bar \sigma + (Z_3-1) (f   \sigma \bar {\cal O} + f \bar \sigma {\cal O} )\,.
\ee 
The  one-particle irreducible structures 
 that may contain divergences
 are $\Gamma_{\sigma \bar \sigma}$, $\Gamma_{\sigma \phi \phi}$ and
 $\Gamma_{\phi \phi \phi \phi}$. The quartic vertex $\Gamma_{\phi \phi \phi \phi}$ 
 is in fact
   subleading in the large $N$ limit,
as illustrated in Figure \ref{subleading} in a one-loop example.
The leading contributions
to the scalar four-point function contain cuttable $\sigma$ propagators. 
This  is an example of a general fact that we will use repeatedly: 1PI diagrams with internal $\sigma$ propagators
are subleading for large $N$. Indeed, adding internal $\sigma$ lines increases the number
of $\phi$ propagators, which are suppressed by $1/N$. 

\smallskip

The upshot is that while for finite $N$   (\ref{Ltree}) is not renormalizable as written (we need to add an explicit $\cO \bO$ counterterm), for large $N$ it is.

From the Feynman rules, we immediately find 
\begin{eqnarray} \label{Gammass}
\Gamma_{\sigma \bar \sigma} (x, y)  &  = & 
 f Z_2 \, \delta(x-y) + Z_3^2 \, f^2 \, \langle\, {\cal O} (x)
\bar {\cal O} (y) \,\rangle_{f=0}  \; ,   \\
\label{Gammaspp}
 \nonumber \\
\Gamma_{\sigma \phi \phi} (x ; y,z ) & = &   -f \, Z_3 \, \langle \,  {\cal O}(x) \phi(y)   \phi(z)  \, \rangle_{f=0}^{1PI} \;.
\end{eqnarray}
Since we are assuming for now that the single-trace action is free, the $f=0$ correlators
appearing above are given by their tree-level expressions.
The three-point function $ \langle  {\cal O} \phi    \phi   \rangle_{f=0}^{1PI} $
 is simply a  constant,  
 \be
 \Gamma_{\sigma \phi \phi} = - f Z_3 \cdot {\rm const} \,.
 \ee
 Clearly no renormalization of the $\sigma \phi \phi$ vertex
 is needed and we can set $Z_3 =1$. On the other hand, the two-point function
  \be
 \langle {\cal O} (x)
\bar {\cal O} (0) \rangle_{f=0} \equiv \frac{v}{2 \pi^2 x^4  }
 \ee
 \smallskip
requires renormalization, since its short-distance behavior is too singular to admit
a Fourier transform.  We adopt the  elegant scheme of differential renormalization \cite{Freedman:1991tk, Freedman:1992gr}.
The singularity is regulated by  smearing the scalar propagator,
\be
 \langle {\cal O} (x)
 \bar {\cal O} (0) \rangle_{f=0}  =   \frac{v}{2 \pi^2}\frac{1}{ ( x^2 + \epsilon^2 )^2} \, ,
  \ee
   where $\epsilon$ is a short distance cutoff. Introducing a dimensionful constant $\mu$, one may separate out
   the divergence as follows,
   \be
   \frac{v}{2 \pi^2}\frac{1}{ ( x^2 + \epsilon^2 )^2}   \stackrel{\epsilon \to 0} {\longrightarrow} - \frac{v}{8 \pi^2} \Box \frac{\ln x^2 \mu^2 }{x^2}  - v \ln \mu \epsilon \; \delta(x) \,.
  \ee
The first term is the renormalized two-point function: it is finite (Fourier transformable) if one interprets the Laplacian as acting to the left under the integral sign.
The constant $\mu$ plays the role of the renormalization scale. Back in (\ref{Gammass}), we take the $Z$-factors to be
\be
Z_2 = 1+  v f  \log \mu \epsilon \, ,\qquad Z_3 =1 \, ,
\ee
and find the renormalized correlator
\be
\Gamma_{\sigma \bar \sigma} (x, y) = f \delta(x-y)   -\frac{v f^2}{8 \pi^2} \Box \frac{\ln  \mu^2 (x-y)^2}{(x-y)^2} \,.
\ee
We are  now in the position to calculate  $\beta_f$ and the anomalous dimension $\gamma_{\cal O}$ of the single-trace operator.\footnote{
Note that $\gamma_{\cal O} $ coincides with $\gamma_\sigma$, since connected correlation functions of $\sigma$ are equal
(for separated points) to connected correlation functions of ${\cal O}$.}
The renormalized two-point function satisfies the Callan-Symanzik equation
\be
\left[ \mu \frac{\partial}{\partial \mu} + \beta_f \frac{\partial}{\partial f} - 2 \gamma_{\cal O} \right ]    \Gamma_{\sigma \bar \sigma}= 0 \,.
\ee
Recalling the identity
\be
\mu \frac{\partial}{\partial \mu} \left[ -\frac{1}{8 \pi^2} \Box \frac{\ln  \mu^2 x^2}{x^2} \right] = \delta(x) \, ,
\ee
we see that the CS equation implies
\begin{eqnarray} \label{linear}
2 f \beta_f - 2 \gamma_{\cal O} f^2   & = &  0 \\
\beta_f - 2 \gamma_{\cal O} f + v f^2 &  = & 0\, ,
\end{eqnarray}
the first condition arising for $x \neq y$ and the second  from the delta function term. Incidentally, 
 the CS equation for $\Gamma_{\sigma \phi \phi}$, namely
\be
\left[ \mu \frac{\partial}{\partial \mu} + \beta_f \frac{\partial}{\partial f} - \gamma_{\cal O} - 2 \gamma_\phi \right ]    \Gamma_{\sigma \phi \phi}= 0\,, \quad \gamma_\phi = 0  \, ,
\ee
immediately gives $\beta_f = f \gamma_\cO$,
equivalent to   (\ref{linear}).
Solving the linear system, we  find
\be
\beta_f = v f^2 \,, \qquad \gamma_{\cal O} = v f \,. 
\ee
These are exact results (all orders in $f$) in the large $N$ theory. The essential point, borne out  by the auxiliary field trick,
is that the for $\lambda = 0$ the only primitively divergent diagram 
is the  one-loop renormalization of the $\sigma$ propagator.

\subsubsection{The general case}

As we take $\lambda \neq 0$, we face the complication that the version of
the theory with the auxiliary field, equation (\ref{Ltree}), is not renormalizable
as it stands, since an explicit quartic term $ \cO \bO$ is regenerated 
by the interactions.  We are led to consider the two-parameter theory
 \be \label{L2}
 {\cal L}^{(2)} (g, h) \equiv  {\cal L}_{ST}  -g \sigma \bar \sigma + g \sigma \bar {\cal O} + 
 g \bar \sigma {\cal O}  + h \cO \bO \,.
 \ee
Comparing with the
 original form of the lagrangian without auxiliary field,
 \be \label{L1}
  {\cal L}^{(1)} (f) \equiv {\cal L}_{ST}  + f \cO \bO \, ,
 \ee
we have the equivalence
\be
{\cal L}^{(1)} ( g + h) \sim {\cal L}^{(2)} (g, h) \,.
\ee
(We  leave implicit the dependence of ${\cal L}^{(1)}$ and ${\cal L}^{(2)}$
on the single-trace couplings $\lambda$ and on $N$.)  
Clearly,
 \be
\beta_f (g +h) = \beta_g(g,h) + \beta_h (g,h) \, ,
\ee
where $\beta_f$ is the beta function for the coupling $f$ in theory (\ref{L1}), and $\beta_g$ and $\beta_h$
are the beta functions for the couplings $g$ and $h$ in theory (\ref{L2}). 
It may appear that not much is gained by considering the more complicated lagrangian  ${\cal L}^{(2)}(g, h)$,
but in fact the auxiliary field trick still provides a useful reorganization of large $N$ diagrammatics.
Our strategy is to work in the theory defined by ${\cal L}^{(2)} (g, h)$, but {\it  in the limit that the renormalized quartic coupling $h \to 0$}. 

\smallskip

 We need not discuss explicitly
 the renormalization of the single-trace part of the action. For  large $N$,
 the 1PI diagrams that renormalize the couplings in ${\cal L}_{ST}(\lambda)$
 are independent of $g$, because leading diagrams at large $N$ do not contain internal $\sigma$ lines. 
  Since we are also taking  $h \to 0$, this implies that the renormalization of ${\cal L}_{ST} (\lambda)$
 proceeds independently of ${\cal L}^{(2)}_{DT}$. We recall that by assumption,
 ${\cal L}_{ST} (\lambda)$ is such that $\beta_\lambda = 0$ for large $N$.

\smallskip

 To discuss the renormalization of ${\cal L}_{DT}^{(2)} (g, h \to 0)$,
we parametrize the  counterterms as 
\be \label{L2ct}
\delta {\cal L}^{(2)}_{DT} =   - (Z_2 -1) g \sigma \bar \sigma + (Z_3-1) (g   \sigma \bar {\cal O} + g \bar \sigma {\cal O} ) + (Z_4-1) h \cO \bO \,.
\ee
As we have emphasized, even for $h \to 0$ a quartic
counterterm $(Z_4-1) h \cO \bO$ is needed in order to 
 cancel the divergence of $\Gamma_{\phi \phi \phi \phi}$. We can use again the fact
 that for large $N$,    $\Gamma_{\phi \phi \phi \phi}$  
 is independent of $g$ (recall Figure \ref{subleading}). Hence for  $h \to 0$ the quartic counterterm can
  only depend on the single-trace coupling $\lambda$,
\be
\lim_{h \to 0} (Z_4-1) h = f( \lambda, \epsilon, \mu) \,.
\ee
It follows that the  corresponding beta function is only a function of $\lambda$,
\be
 \beta_h (g, h = 0) = a (\lambda) \, .  \qquad
\ee
In orbifolds of ${\cal N}=4$ SYM, $\lambda$ is the usual 't Hooft coupling,
and $a(\lambda)$ has a perturbative expansion of the form
\be \label{pert1}
a(\lambda) = \sum_{L=1}^\infty a^{(L)} \lambda^{L+1} \, ,
\ee
where $L$ is the number of loops.

\smallskip
The analysis of the two remaining primitively divergent structures, $\Gamma_{\sigma \bar \sigma}$
and $\Gamma_{\sigma \phi \phi}$, proceeds similarly as in the $\lambda = 0$ case, with
a few extra elements.
We have (for $h=0$),
\begin{eqnarray} \label{Gammass2}
\Gamma_{\sigma \bar \sigma} (x, y)  &  = & 
 g Z_2 \, \delta(x-y) + Z_3^2 \, g^2 \, \langle\, {\cal O} (x)
 \bar {\cal O} (y) \,\rangle_{g=h=0}  \; ,   \\
\label{Gammaspp2}
 \nonumber \\
\Gamma_{\sigma \phi \phi} (x ; y,z ) & = &   -g \, Z_3 \, \langle \,  {\cal O}(x) \phi(y)   \phi(z)  \, \rangle_{g=h=0}^{1PI} \;.
\end{eqnarray}
From the last equation, we see that the factor $Z_3$ has the role of renormalizing the composite operator  ${\cal O}$
in the theory with $g=h=0$,
\be
{\cal O}_{g=h=0}^{ren} \equiv Z_3 (\lambda, \mu, \epsilon) \, {\cal O}    \,.
\ee
The  dependence of ${\cal O}_{g=h=0}^{ren} $ on the renormalization scale $\mu$ is given by 
\be
\mu \frac{\partial}{\partial \mu} \, {\cal O}_{g=h=0}^{ren}  =- \gamma(\lambda)\, {\cal O}_{g=h=0}^{ren} \, ,
\ee
where $\gamma (\lambda)$ is, by definition,  the anomalous dimension of the single-trace operator 
 in the theory where we set to zero the double-trace couplings. The two-point function of ${\cal O}^{ren}_{g=h=0}$ takes then the standard form
   \be \label{naive}
\langle \cO^{ren} (x) \cO^{ren}(0) \rangle_{g=h=0} = \frac{v(\lambda)  }{2 \pi^2} \frac{\mu^{-2 \gamma(\lambda)}}{ x^{4 + 2 \gamma(\lambda)} }\, , \quad x \neq 0 \, .
\ee
We have indicated that the normalization $v$ will in general depend on $\lambda$. In orbifolds of ${\cal N}=4$,
$v(\lambda)$ and $\gamma(\lambda)$ have perturbative expansions of the form
\be \label{pert2}
v(\lambda) = \sum_{L=1}^\infty v^{(L)} \lambda^{L-1} \, ,\qquad \gamma(\lambda) = \sum_{L=1}^\infty \gamma^{(L)} \lambda^{L} \,.
\ee
The expression (\ref{naive}) is not well-defined  at short distance and needs further renormalization, which we perform again
in the differential renormalization scheme.
We first expand
\be
\frac{\mu^{-2 \gamma}}{ x^{4 + 2 \gamma } } = \sum_{n=0}^\infty  \frac{(-\gamma )^n}{n!} \,\frac{\log^n \mu^2 x^2}{x^4} \ ,
\ee
and then renormalize each term of the series using the substitutions \cite{Latorre:1993xh}
\be
\frac{\log^n \mu^2 x^2 }{x^4} = -\frac{n!\,}{4} \; \Box \sum_{k=1}^{n+1} \frac{1}{k!} \, \frac{\log^k \mu^2 x^2}{x^2} \, .
\ee
These are exact identities for $x\neq 0$ and provide the required modification of the behavior at $x=0$, if one stipulates
that free integration by parts is allowed under the integral sign.

\smallskip

Back in (\ref{Gammass2}), we have\footnote{The value of $Z_2$ is defined implicitly by this equation. As in the $\lambda = 0$ case, we could 
introduce a short-distance cutoff $\epsilon$ and then choose $Z_2 (\epsilon, \mu)$ such that  the final result (\ref{Gammass3}) for the fully renormalized
correlator is obtained.}
\begin{eqnarray} 
 \Gamma_{\sigma \bar \sigma} (x, 0)      &  =  & 
 g Z_2 \, \delta(x) +  g^2 \, \langle\, {\cal O}^{ren} (x)
 {\cal O}^{ren} (0) \,\rangle_{g=h=0}  
\\  \label{Gammass3}
& =  & g\,\delta (x) -g^2   \frac{v }{8 \pi^2}  \sum_{n=0}^\infty  (-\gamma)^n\, \Box  \, \sum_{k=1}^{n+1}  \frac{1}{k!} \, \frac{\log^k (\mu^2 x^2)}{x^2}\, .
\end{eqnarray}
The CS equation,
\be \label{CSg}
\left[ \mu \frac{\partial}{\partial \mu} + \beta_g \frac{\partial}{\partial g} - 2 \gamma_\cO \right ]    \Gamma_{\sigma \bar \sigma}= 0 \,,
\ee
gives as before two conditions, one for $ x \neq 0$ and one from the delta function term. For $ x \neq 0$,
we may simply use the naive expression (\ref{naive}) for the correlator, and we find
\be \label{linear1}
2 g \beta_g - 2 \gamma_\cO g^2 - 2 \gamma  g^2= 0 \,.
\ee
It is easy to check that the same condition follows from the CS for $\Gamma_{\sigma \phi \phi}$.
On the other hand, terms proportional to $\delta(x)$  in (\ref{CSg}) arise either from the explicit $g \delta(x)$ in $\Gamma_{\sigma \bar \sigma}$,
or  when the $\mu$ derivative hits the $k=1$ terms of the series,
\be  \label{linear2}
0 = \beta_g - 2 \gamma_\cO g + g^2 v \sum_{n=0}^{\infty} (-\gamma )^n=   \beta_g - 2 \gamma_\cO g + \frac{g^2 v}{1 + \gamma} \,.
\ee
The solution of the linear system (\ref{linear1}, \ref{linear2}) is
\be 
\gamma_\cO =\gamma +\frac{v g}{1 + \gamma }  \, , \quad \beta_g =\frac{ v g^2 }{1 + \gamma}+ 2 g \gamma \,.
\ee
We can finally evaluate $\beta_f$  in the original theory (\ref{L1}). From
\be
\beta_f (f)= \beta_g(g=f,h=0) + \beta_h (f,h=0) \, ,
\ee
we find
\be \label{betaf}
\boxed{  \phantom{ \Bigg( }
\beta_f = \frac{v (\lambda)}{1+ \gamma(\lambda)} \, f^2 + 2  \, \gamma(\lambda)\,  f + a(\lambda) \, .
} 
\ee
This is the sought generalization of the one-loop result (\ref{betaf1}) originally found in \cite{Dymarsky:2005uh}.  The expression for the full conformal dimension of the single-trace operator is
\be \label{DeltaO}
 \boxed{ \phantom{\Bigg( } 
\Delta_\cO = 2 + \gamma_{\cal O} (f, \lambda) = 2 + \gamma(\lambda) + \frac{v(\lambda)}{1 + \gamma(\lambda) } \, f \,.
}
\ee
The boxed equations are valid to all orders in large $N$ perturbation theory.

\section{  Double-trace running and dynamical symmetry breaking}

\label{CW}

The beta function of the double-trace coupling remains quadratic in $f$,
to all orders in planar perturbation theory. This simplification
allows to draw some general conclusions about the behavior of the running coupling
and  the  stability of the Coleman-Weinberg potential. While the essential physics
is already visible in the one-loop approximation, it seems worthwhile to pursue
a general analysis.

\subsection{Running coupling}

\label{running}

 We need to distinguish two cases,
according to whether the quadratic equation
\be \label{beta0}
\beta_f =  \frac{v (\lambda)}{1+ \gamma(\lambda)} \, f^2 + 2  \, \gamma(\lambda)\,  f + a(\lambda)= 0 
\ee
has  real or complex zeros. We define the discriminant $D(\lambda)$, 
\be
D (\lambda) \equiv \gamma(\lambda)^2 -  \frac{a(\lambda) v(\lambda)}{1 + \gamma(\lambda) } \, ,
\ee
 and the square root of $|D|$,
 \be \label{bpert}
 b (\lambda)\equiv \sqrt{|D(\lambda)|} \,.
 \ee
From (\ref{pert1}, \ref{pert2}),  $b(\lambda)$ has a perturbative expansion of the form
 \be
 b(\lambda) = b^{(1)} \lambda + b^{(2)} \lambda^2 + \dots \,.
 \ee

 \bigskip
 
\noindent $\bullet$ {\it Positive discriminant}
 
 \bigskip
 
   \begin{figure}[t] 
  \centering 
\mbox{\subfigure[$D>0$]{\epsfig{figure=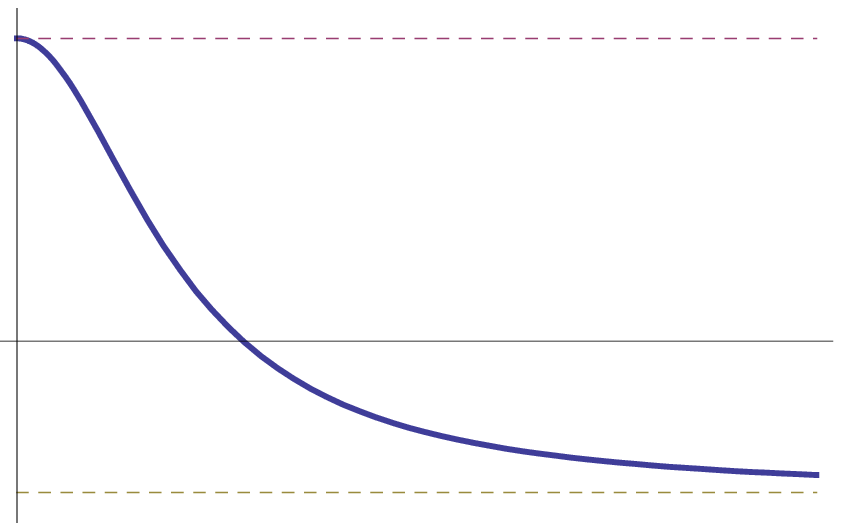,width=6cm,angle=0}}\quad  \quad \quad  \quad  \quad  \quad  \quad \quad
\subfigure[$D<0$]{\epsfig{figure=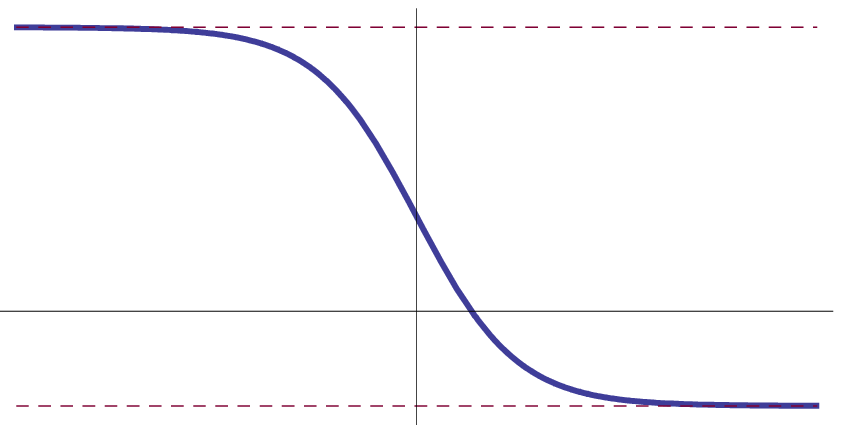,width=5cm,angle=90}}  } 
      \put(-350,95){$f_+$}     
            \put(-350,0){$f_-$}     
            \put(-165,30){$\mu$}  
   \put(-95,10){$\mu_{IR}$}   
    \put(0,130){$\mu_{UV}$}   
      \put(-55,100){$f(\mu)$}   
        \put(-250,70){$f(\mu)$}   
 \caption{ \it The two qualitative behaviors of the running coupling $f(\mu)$ for $D >0$ and $D < 0$.} 
  \label{flow} 
\end{figure}
 
  If $D > 0$, (\ref{beta0}) has real solutions
 \be \label{fpm}
 f_{\pm} = -\frac{\gamma}{\tilde v} \pm \frac{b }{\tilde v} \, , \quad \tilde v \equiv \frac{v}{1 + \gamma} \,.
 \ee
 In this case we can maintain conformal invariance in the quantum theory by tuning
 $f$ to one of the two fixed points.
Since $v>0$ (the two-point function of ${\cal O}$ is positive by unitarity), we see that
$f_-$ is  UV stable and $f_+$  IR stable. The differential equation for the running coupling,
\be \label{df}
\mu \frac{\partial}{\partial \mu} f(\mu) = \beta_f (f(\mu)) \, ,
\ee
is easily solved to give
\be
 f (\mu) = \frac{\left( \frac{\mu}{\mu_0} \right)^{2 b} f_- + f_+ }{\left( \frac{\mu}{\mu_0} \right)^{2 b}  +1   } \,.
\ee
The function $f(\mu)$ is plotted on the left in Figure \ref{flow}. The running coupling interpolates
smoothly between the IR and the UV fixed points.

\bigskip
\noindent $\bullet$ {\it Negative discriminant}
 
 \bigskip

If  $D < 0$ there are no fixed points for real $f$ and conformal invariance is broken in the quantum theory. 
The solution of (\ref{df}) is
\be
 f(\mu) =  -\frac{\gamma}{\tilde v } +\frac{ b} {\tilde v} \tan \left[ \frac{b}{\tilde v} \ln (\mu/\mu_0)  \right] \, , \quad \tilde v \equiv \frac{v}{1+\gamma}\,.
\ee
There are Landau poles both in the UV and in the IR, at
energies 
\begin{eqnarray}
\mu_{IR}  & = &  \mu_0 \, \exp\left(-\frac{\pi \tilde v}{2 b} \right)  \cong   \mu_0 \, \exp\left(-\frac{\pi v^{(1)}}{2 b^{(1)} \lambda} \right)   \, 
\\
\mu_{UV}  & = &  \mu_0 \, \exp \left( \frac{\pi \tilde v}{2 b} \right) 
 \cong \mu_0 \, \exp\left(\frac{\pi v^{(1)}}{ 2 b^{(1)} \lambda} \right)  \,.
\end{eqnarray}
The behavior of $f(\mu)$ is plotted on the right in Figure \ref{flow}. 

\subsection{Effective potential}

The running of the double-trace coupling $f$ and the
generation of a quantum effective potential for the scalar fields
are closely related. We wish to make this relation precise.

\smallskip

Let us consider a spacetime independent vev for the scalars,
\be
\langle\,  \phi_a^{i\; b} \, \rangle = \varphi \; T_a^{i \;b} \, .
\ee
We have picked some direction in field space specified by the tensor $T_a^{i \;b}$, where $i$ is a flavor index and $a, b=1, \dots N$ are color indices.
We need not assume that it is a classical flat direction. With no loss of generality
 we take $\varphi  \geq 0$. 

\smallskip
We now go through the textbook renormalization group analysis of 
 the quantum effective potential ${\cal V}(\varphi)$.  The RG equation reads
\be \label{RG}
\left[  \mu \frac{\partial}{\partial \mu} + \beta_f \frac{\partial}{\partial f}  - \gamma_\phi \, \varphi \frac{\partial}{\partial \varphi} \right]  {\cal V}(\varphi, \mu, f, \lambda) = 0 \, ,
\ee
where $\gamma_\phi (\lambda)$ is the anomalous dimension of the scalar field $\phi$. Note
that for large $N$, $\gamma_\phi (\lambda)$ is independent of $f$.  Writing (\ref{RG}) as
\be
 {\cal V}(\varphi, \mu, f, \lambda) \equiv \varphi^4 \, U(\varphi/\mu, f, \lambda) \, , \quad
\left[ \varphi \frac{\partial}{\partial \varphi} - \frac{\beta_f}{1 + \gamma_\phi} \frac{\partial}{\partial f} + \frac{4 \gamma_\phi}{1+ \gamma_\phi} \right]  U = 0 \,  ,
\ee
one finds that  the most general solution takes the form
\be
{\cal V} (\varphi, \mu, f, \lambda) = \varphi^4  \,  \left( \frac{\varphi}{\mu}\right)^{- \frac{4 \gamma_\phi}{1+ \gamma_\phi}}  \, U_0 (\hat f(\varphi), \lambda)   \, ,
\ee
where $\hat f (\mu)$ satisfies
\be \label{dfb}
 \mu \frac{\partial}{\partial \mu} \,  \hat f (\mu) = \frac{\beta_f (\hat f) }{  1 + \gamma_\phi} \, .
\ee
In general, the arbitrary function $U_0 (\hat f, \lambda)$ is found order by order by comparing with explicit perturbative results.
In our case, because of large $N$, the  double-trace coupling contributes to the effective potential  only at tree-level.
This is again a consequence of the fact that 1PI diagrams with internal $\sigma$ lines are suppressed. 
Moreover, by assumption the  single-trace
quartic term $N \lambda {\rm Tr} \phi^4$ is not renormalized at large $N$,
so that the {\it explicit}  $\lambda$ dependence of $U_0 (\hat f, \lambda)$ is also exhausted
by the tree-level contribution. There is of  course an implicit $\lambda$ dependence in $\hat f$, as clear from (\ref{dfb}, \ref{betaf}).
The full tree-level
contribution to the effective potential is
\be \label{Vtree}
{\cal V}_{tree} (\varphi) =  N \lambda {\rm Tr } \phi^4 + f \, \cO \bar {\cal O}  =  N^2 ( C_{ST}  \lambda    +  C_{DT}  f  ) \, \varphi^4 \, ,
\ee
where $C_{ST}$ and $C_{DT}$ are some
 non-negative proportionality constants of order one.\footnote{We are suppressing flavor indices: $N \lambda \Tr \phi^4$ in (\ref{Vtree}) is
 a shortcut for the scalar potential of the single-trace lagrangian ${\cal L}_{ST}$, which we require to be bounded from below.
 Then $C_{ST} \geq 0$. On the other hand,  positivity of $C_{DT}$ is clear from (\ref{Vtree}),
since $ \cO \bO$ is a positive quantity.} If the vev is taken along a classical flat direction of the single-trace lagrangian, then $C_{ST} = 0$,
but we need not assume this is the case. Thus 
 \be
 U_0( \hat f, \lambda) =  N^2 ( C_{ST}  \lambda    +  C_{DT} \hat f  ) \, .
 \ee
 The final result for the large $N$ effective potential is
 \be \label{V}
 \boxed{ \phantom{\Bigg( } 
{\cal V} (\varphi) = N^2  \, \mu^{\frac{4 \gamma_\phi }{1+ \gamma_\phi}} \, \left[                C_{ST} \, \lambda    +  C_{DT} \, \hat f  (\varphi)            \right]  \varphi^{\frac{4 }{1+ \gamma_\phi}}  \,  \,.
}
\ee
Ordinarily,  at a fixed order in perturbation theory the RG improved effective potential 
  can be trusted in the range of $\varphi$ such that the running coupling $\hat f (\varphi)$ is small.
 In our case,   ${\cal V}(\varphi)$ receives no higher corrections in $\hat f$, so it appears that
   (\ref{V}),    being the full non-perturbative answer,  may have a broader validity.

\smallskip

Let us make contact with the explicit one-loop expression of the effective potential. To this order, 
\be \label{to}
\tilde v(\lambda) \cong v^{(1)}\, , \quad \,   \gamma(\lambda)  \cong \gamma^{(1)} \lambda \,, \quad a(\lambda) \cong a^{(1)} \lambda^2 \, ,\quad  \gamma_\phi  \cong \gamma_{\phi }^{(1)} \lambda \, ,
\ee
and the expansion of (\ref{V}) gives
\begin{eqnarray} \label{V1}
&& {\cal V}_{1-loop} (\varphi)  =\\
&&  N^2 \varphi^4 \log\left(\frac{\varphi}{ \mu}\right) \cdot   \left[   \,  v^{(1)} f^2 C_{DT}+ 2 f \lambda ( \gamma^{(1)} - 2 \gamma_{\phi }^{(1)}  )C_{DT} + \lambda^2 (a^{(1)} C_{DT} - 4 \gamma_{\phi }^{(1)} C_{ST} ) \,  \right] \, .\nonumber
\end{eqnarray}
Each term has an obvious diagrammatic interpretation.

\subsection{Stability versus conformal invariance }

\label{versus}

Armed with the general form (\ref{V}) of the large $N$ effective potential, we can investigate
the stability of the symmetric vacuum at $\varphi = 0$. Since the single-trace coupling $\lambda$ does not run, we can treat it as an external
 parameter.  For given $\lambda$,  the functions $a(\lambda)$, $\tilde v(\lambda)$, $\gamma(\lambda)$ and $\gamma_\phi(\lambda)$
 are just constant parameters that enter the expression for ${\cal V} (\varphi)$.
 
 \smallskip

The qualitative behavior of ${\cal V}( \varphi)$ is dictated by the discriminant $D(\lambda)$.
Comparing (\ref{dfb}) with (\ref{df}),  we see that $\hat f (\varphi)$ behaves just as  $f(\varphi)$, 
up to some trivial rescaling of coefficients by $1/(1+ \gamma_\phi)$. 
We consider again the two cases:

\bigskip
\noindent $\bullet$ {\it Positive discriminant}
 
 \bigskip
 
 For $D >0$, the running coupling is given by
 \be \label{bfrun}
 \hat f (\varphi) = \frac{\left( \frac{\varphi}{\mu} \right)^{2 \hat b} f_- + f_+ }{\left( \frac{\varphi}{\mu} \right)^{2  \hat b}  +1   } \, , 
 \qquad \hat b  \equiv  \frac{b  }{1 + \gamma_\phi }\,.
 \ee
The constant solutions $\hat f (\varphi) = f_{\pm}$ are obtained as degenerate cases for $\mu \to 0 $ and $\mu \to \infty$. 
In the generic case, the effective potential is bounded
by  the two functions (we set $\mu \equiv 1$)
\be
N^2\, ( C_{ST} \lambda +C_{DT}  f_+ )\,  \varphi^{\frac{4 }{1+ \gamma_\phi}}   \leq  \,  {\cal V}(\varphi) \, \leq N^2\, ( C_{ST} \lambda + C_{DT}  f_- )\,  \varphi^{\frac{4 }{1+ \gamma_\phi}}  \, ,
\ee
where the lower bound is attained for $\varphi \to 0$ and the upper bound for $\varphi \to \infty$. Recall from (\ref{fpm})
that $f_- < f_+$, with $f_-$ always negative.  If
\be \label{fpp}
 C_{ST} \lambda + C_{DT} f_{+}  > 0 \, ,
 \ee 
 then $\varphi = 0$ is at least a local minimum, otherwise it is a global maximum and the potential is unbounded from below.
 Condition (\ref{fpp}) is simply the requirement that the tree-level potential (\ref{Vtree}) be bounded from below when $f$ is set to its IR fixed point $f_+$.  If (\ref{fpp}) holds, it is also permissible
 to simply pick the constant solution $\hat f (\varphi) = f_+$. Then ${\cal V}$ is monotonically increasing and $\varphi = 0$ is the global minimum.
 In the generic case (\ref{bfrun}), we need the stronger condition
\be C_{ST} \lambda + C_{DT} f_{-}  > 0
 \ee 
 to ensure that the potential is bounded from below. Then 
  $\varphi = 0$ is the global minimum. 
  
  \smallskip

 In view of the comments below (\ref{V}), we believe that this analysis has general validity. It is certainly
 valid for $\lambda \ll 1$, since then
 $f_{\pm} \sim \lambda + O(\lambda^2)$,  and the effective coupling $\hat f(\varphi) \ll1 $  for every value of $\varphi$. 
 
  \smallskip
 In summary,  barring  pathological
 cases where the potential is unbounded from below, for $D > 0$ the vacuum $\varphi = 0$ is stable
 and dynamical symmetry breaking does not occur.

\bigskip
\noindent $\bullet$ {\it Negative discriminant}
 
 \bigskip

If $D <0$, the effective potential reads, in units $\mu \equiv 1$,
\be
{\cal V}(\varphi) =  N^2 \left[ C_{ST} \lambda  + C_{DT} \hat f(\varphi) \right]  \varphi^{\frac{4}{1+ \gamma_\phi} }\, , \quad \hat f(\varphi)=  - \frac{\gamma}{\tilde v} + \frac{b}{\tilde v}  \, \tan \left( \frac{b}{\tilde v}  \log \varphi \right) \,.
\ee
The theory only makes sense as an effective field theory
for energy scales intermediate between the two Landau poles, $\mu_{IR} = e^{-\frac{\pi}{2 b}} \ll \varphi \ll \mu_{UV}=  e^{+\frac{\pi}{2 b}}$.
The potential ranges between minus infinity at $\mu_{IR}$ and plus infinity at $\mu_{UV}$.
A little algebra shows that ${\cal V}(\varphi)$ is either a monotonically increasing function,
or it admits a local maximum and a local minimum. Local extrema exist if
\be \label{mincond}
 \lambda\, \frac{C_{ST}}{C_{DT}} - \frac{\gamma}{\tilde v} <  \frac{1}{1 + \gamma_\phi} - \frac{b^2 (1 + \gamma_\phi)}{4 {\tilde v}^2}  \, ,
\ee
with the potential    always negative at the local minimum,
\be \label{Vneg}
{\cal V } (\varphi_{min}) < 0 \, .
\ee
From (\ref{to}, \ref{bpert}), we see that  (\ref{mincond})
 is always obeyed for sufficiently small $\lambda$. The value of the running coupling
at the minimum can be expanded for $\lambda \ll 1$,
\begin{eqnarray} \label{fst}
 \hat f (\varphi_{min})  = -\alpha \, \lambda +\left( - \frac{\; \;a^{(1)}}{4} + \gamma_\phi^{(1)}  \, \alpha -\frac{\; v^{(1)}}{4}\, \alpha^2 \right) \lambda^2 + O(\lambda^3) \,,\quad \alpha \equiv \frac{C_{ST}}{C_{DT}} \,.
 \end{eqnarray}
For small $\lambda$, $\hat f (\varphi_{min})$ is also small,  the local
minimum can  be trusted, and dynamical symmetry breaking occurs. 
If the vev is taken along a flat direction for the single-trace potential, namely if $C_{ST} = 0$, then
the double-trace coupling at the new vacuum is of order $O(\lambda^2)$, which is perhaps the more
familiar behavior -- as in the original analysis of massless scalar electrodynamics \cite{Coleman:1973jx}. From (\ref{mincond}, \ref{Vneg}, \ref{fst}), we find that for 
small  $\lambda$ symmetry breaking occurs even if the  tree level single-trace potential does not vanish $(C_{ST} \neq 0)$.

\smallskip

We take the liberty to belabor this  conclusion, giving an alternative derivation.
One can first expand the effective potential to lowest non-trivial order,
\be
{\cal V}(\varphi) \cong N^2 [ C_{ST} \lambda + C_{DT} \hat f (\mu) ] + {\cal V}_{1-loop} (\varphi) \,  ,
\ee
with ${\cal V}_{1-loop}   $  given by (\ref{V1}).  In looking for the minimum,  ${\cal V}'(\varphi_{min}) = 0$, ${\cal V}''(\varphi_{min}) > 0$,
it is convenient to set the renormalization scale $\mu \equiv \varphi_{min}$. 
Then we just solve for   $\hat f (\varphi_{min})$ and easily reproduce (\ref{fst}).
This is a consistent procedure provided we can {\it find} a renormalization trajectory
where $\hat f(\varphi_{min})$ takes the value (\ref{fst}). A glance at Figure \ref{flow}
shows that yes, we can set $\hat f$ to any prescribed value. 
Finally, since (\ref{fst}) happens to be small  for $\lambda $ small, the whole analysis can be trusted in
 perturbation theory.

\smallskip

The inequality (\ref{mincond})  can be satisfied also if $\lambda$ is of order one, in which case $\hat f (\varphi)$ is of order one.
In view of our remarks about the non-perturbative validity of ${\cal V}(\varphi)$, it seems plausible
that the local minimum can also be trusted in this case.

\section{AdS/CFT }

\label{prescription}

We have used standard field theory arguments to characterize the two possible
behaviors for a large $N$ theory conformal in its single-trace sector.
Either all double-trace beta functions admit real zeros,  and then
the symmetric vacuum is stable and conformal invariance is preserved;
or at least one beta function has no real solutions, and then
conformal invariance is broken and dynamical
symmetry breaking occurs.

\smallskip
We now give a reinterpretation  of these results in  light of the AdS/CFT correspondence.
Even for negative discriminant, we insist in solving for the zeros
of the double-trace beta function,
\be
f_{\pm} = -\frac{\gamma}{\tilde v} \pm \frac{ \sqrt{D } }{\tilde v} \, . \quad 
\ee
Setting $f  = f_{\pm}$,  the full conformal dimension (\ref{DeltaO}) of the single-trace operator ${\cal O}$ reads
\be
\Delta_{\cal O} = 2 + \gamma + \tilde v f_\pm = 2 +  { \gamma }- \gamma  \pm \sqrt{D}  = 2 \pm \sqrt{D} \,.
\ee
 So at the fixed point, the anomalous dimension of ${\cal O}$ is either real if $D >0$
   or {\it purely imaginary} if $D < 0$.
 This is just as expected from the AdS/CFT formula
 \be
 \Delta_{\cal O} = \frac{d}{2} \pm \sqrt{\frac{d^2}{4} + m^2 R^2} = 2 \pm \sqrt{4 + m^2 R^2} \, ,
 \ee
where $m$ is the mass of the $AdS_5$ scalar field dual to ${\cal O}$, if we identify
\be \label{m2}
\boxed{   \phantom{ \Bigg( }
m^2 (\lambda) R^2 = m^2_{BF} R^2  + D(\lambda)  = - 4 + D(\lambda)   \, .
}
\ee
For $D >0$, we are in the standard situation of real coupling constant, real anomalous dimension
and  dual scalar mass  above the stability bound, $m^2 > m^2_{BF}$. 
We propose to take (\ref{m2}) at face value even when {$D <0$}. 
If $m^2 < m^2_{BF}$, the AdS bulk vacuum is unstable. Similarly, if $D < 0$, the field theory conformal-invariant vacuum is unstable. 
 Equation (\ref{m2}) gives the precise relation between the two instabilities.
The proper treatment of both the bulk and the boundary theory would be to expand around the stable minimum.
But in stating that the AdS scalar has a certain mass $m^2 < m^2_{BF}$, we are implicitly
quantizing the bulk theory in an AdS invariant way.
The dual statement is to formally quantize the boundary theory in a conformal invariant way, 
around the symmetric minimum $\varphi = 0$, by tuning the coupling to
the complex fixed point $f = f_+$ (or $f_-$). At either fixed point, the operator dimension is complex,
 \be
 \Delta_{\cal O} = 2 \pm  i \; b \, .
\ee

\smallskip

The discriminant $D(\lambda)  =\gamma(\lambda)^2 - a (\lambda)\tilde v(\lambda) $ is a purely field-theoretic quantity. 
In principle (\ref{m2}) is a prescription to compute the tachyon mass from the field theory, 
at least order by order in perturbation theory.
 It would be interesting to see if integrability techniques \cite{Beisert:2005he} are applicable to this problem,
 though the fact that ${\cal O}$ is a ``short'' operator may represent a challenge.
  For now we may
 compare field theory results at weak coupling with the strong coupling behavior
 predicted by the gravity side. Let us look at a couple of examples.

\subsection{Two examples }

\label{examples}

Expanding (\ref{m2}) to one-loop order,
\be
m^2 (\lambda) R^2 =  - 4 +  D(\lambda) = -4 + \left[ ( \gamma^{(1)})^2 - a^{(1)} v^{(1)}  \right] \lambda^2 + O(\lambda^3) \, .
\ee
The coefficients $v^{(1)}$,  $\gamma^{(1)}$, and $a^{(1)}$ were computed in \cite{Dymarsky:2005uh, Dymarsky:2005nc} for several
 orbifolds of ${\cal N} = 4$ SYM.  Obtaining the corresponding $m^2$ is an exercise in arithmetic.

\smallskip

As a first illustration, take
 the $\mathbb{Z}_2$ orbifold theory that arises on  a stack of $N$ electric and $N$ magnetic D3 branes of Type 0B string theory. There
 are twisted scalars in the $\bf 20'$ and $\bf 1$ representations of $SU(4)_R$. From the results in \cite{Tseytlin:1999ii, Dymarsky:2005uh}, one finds
\be
m_{\bf 20'}^2  R^2  \cong  -4  -  \frac{\lambda^2}{8 \pi^4} + O (\lambda^3)  \ ,
\qquad    m_{\bf 1}^2  R^2  \cong   -4  -   \frac{23\lambda^2}{64 \pi^4} + O (\lambda^3)\,.
\ee
Since this  orbifold has fixed points on the $S^5$ (it fixes the whole sphere), 
we expect these masses to  remain negative below the stability bound for all $\lambda$, with the asymptotic behavior
\be
m^2 (\lambda)  R^2  \sim  - \frac{R^2}{\alpha'} = - \lambda^{1/2}\, , \quad \lambda \to \infty \,.
\ee
Let us also consider a simple class of  non-supersymmetric freely acting orbifold,
  $\mathbb{Z}_k$ orbifold with $SU(3)$ global symmetry \cite{Dymarsky:2005uh}. 
  The $\mathbb{Z}_k$ action is
 \be
 z_i \to \omega_k^{\, n} \;  z_i \, ,\quad \omega_k \equiv e^{\frac{2\pi i}{ k} } \, , \quad  n=1, \dots k \, ,
\ee
where $z_i$, $i=1,2,3$ are the three complex coordinates of $\mathbb{R}^6 = \mathbb{C}^3$.
The orbifold is freely acting for $k$ odd, and breaks supersymmetry for $k >3$. Let us focus on the $\mathbb{Z}_5$ case.
There are twisted operators ${\cal O}_{{ \bf 8} , n} $  ${\cal O}_{ {\bf 1} , n} $, with $n=1,2$,   in the octet and singlet of the $SU(3)$ flavor group.
It turns out that in the one-loop approximation the $n=1$ operators have positive discriminant, while the $n=2$ operators
have negative discriminant. From the results of \cite{Dymarsky:2005uh}, one calculates
\be \label{m2expl}
m_{{\bf 8} , 2}^2  R^2  \cong  -4  - \frac{\sqrt{5}-1}{640 \pi^4} \, \lambda^2 + O(\lambda^3) 
 \, ,\quad
    m_{{\bf 1} , 2}^2  R^2  \cong -4 -\frac{7 \sqrt{5} - 1 }{1600 \pi^4} \, \lambda^2  +  O(\lambda^3) \,.
    \ee
The conjectural behavior of $m^2 (\lambda)$ for freely acting orbifolds
is plotted in Figure \ref{proposal}  in the introduction. The one-loop calculation (\ref{m2expl})
gives the second derivative  at $\lambda = 0$. For large $\lambda$,  these
states correspond to highly stretched strings on the $S^5$. The asymptotic behavior 
should thus be
\be
m^2 (\lambda) R^2 \sim \,  \frac{R^4}{\alpha'^2 } \sim \, \lambda \, , \quad \lambda \to \infty \,.
\ee
Figure \ref{proposal} plots the simplest interpolation between the small and large $\lambda$ limits. It would
be very interesting to compute the $O(\lambda^3)$ corrections to (\ref{m2expl}): 
 this picture suggests that they should be positive.

\subsection{Classical flat directions and instability}

The $\mathbb{Z}_{2 k+1}$ freely-acting orbifolds serve as an illustration of another point --
classical flat directions are immaterial in our context. 
The classical moduli space of the theory is $(\mathbb{C}^3/\mathbb{Z}_{2 k+1})^{N}/S_N$. In the brane picture this corresponds to the positions of 
the $N$ D3 branes on the orbifold space $\mathbb{C}^3/\mathbb{Z}_{2k+1}$. 
 The flat directions are parametrized by vevs for the bifundamental scalars  (there are no adjoints). 
Along the flat directions, all twisted operators have zero vev. 

\smallskip

As emphasized in \cite{Adams:2001jb},  this is the case in general
for freely acting orbifolds: they have no adjoint scalars and hence no classical branch along which the  twisted operators could develop a vev. 
However, this does not imply that the symmetric vacuum is stable. On the contrary,
we have seen in section \ref{versus}  that dynamical symmetry breaking occurs
at small coupling  whenever $D <0$, irrespective of the classical potential. Since one can always
find a double-trace coupling with $ D <0$, whether the orbifold is freely acting or not   \cite{Dymarsky:2005nc}, 
we conclude that freely acting orbifolds also have a CW instability which drives into condensation a twisted operator,  $\langle {\cal O} \rangle \neq 0$.
The instability  occurs  away from the flat directions.

\smallskip 

 This reconciles 
 the proposal of \cite{Dymarsky:2005uh}, which relates bulk tachyons with the breaking of conformal invariance, with the general viewpoint
of \cite{Adams:2001jb}, which relates them to the Coleman-Weinberg instability. A detailed analysis of the CW instability in some examples of freely acting orbifolds
has been pursued by \cite{princeton}.

\section{Discussion}

The logarithmic running of double-trace couplings $ f \cO \bO$, where $\cO \sim \Tr \phi^2$, 
is a general feature of large $N$ field theories that contain
scalar fields. In this paper we have studied the renormalization of double-trace couplings
in  theories that have vanishing single-trace beta functions at large $N$. We have derived general
expressions for the double-trace beta function $\beta_f$, the conformal dimension $\Delta_{\cO}$ and
the effective potential ${\cal V}(\varphi)$. The main point is  that $\beta_f$ is a quadratic function of $f$ (and $\Delta_{\cal O}$ a linear
function of $f$), to all-orders in planar perturbation theory, with coefficients that depend on the single-trace couplings $\lambda$.

\smallskip

Double-trace running plays an important role in non-supersymmetric examples of the AdS/CFT correspondence.
We have related the discriminant $D(\lambda)$ of $\beta_f$ to the mass $m^2 (\lambda)$ of the bulk scalar dual to the single-trace operator ${\cal O}$.
If $D (\lambda) <0$, the bulk scalar is a tachyon; on the field theory side, conformal invariance is broken
and dynamical symmetry breaking occurs.  

\smallskip

The authors of \cite{Dymarsky:2005nc} considered orbifolds of ${\cal N}=4$ SYM,
realized as the low energy limit of the theory on $N$ D3 branes at the tip
of the cone $\mathbb{R}^6/\Gamma$. They found a one-to-one
correspondence between double-trace couplings with negative
discriminant and twisted tachyons in the tree-level spectrum  of the type IIB background
before the decoupling limit, namely
$\mathbb{R}^{3,1} \times \mathbb{R}^6/\Gamma$.
(Note that these flat-space tachyons are conceptually distinct from the tachyons in the curved  $AdS_5 \times S^5/\Gamma$ background
that have been the focus of this paper.)\footnote{The  correspondence between twisted sector tachyons and field theory instabilities was first observed
in \cite{Armoni:2003va} in the context of non-commutative field theory.}
It turns out that for {\it all} non-supersymmetric examples in this class,  at least one double-trace coupling has negative
discriminant, and conformal invariance is broken. 

\smallskip

 It will be interesting to investigate more general
constructions to see if conformal examples exist, both
as a question of principle and in view of phenomenological applications.\footnote{See {\it e.g.} \cite{Frampton:1999yb, Frampton:2007fr} for an approach
to conformal phenomenology.}
One possibility, suggested by the correspondence found in \cite{Dymarsky:2005nc},
is to add discrete torsion in a way that removes the tree-level tachyons \cite{Kakushadze:2007qx}. 
Another is to add appropriate orientifold planes.  A promising candidate for a conformal orientifold theory is the $U(N)$ gauge theory with six scalars in the adjoint and
four Dirac fermions in the antisymmetric representation of the gauge group  \cite{Armoni:2007jt}.

\smallskip

Another important question, which is being investigated by \cite{princeton},
is to analyze the IR fate of non-supersymmetric 
orbifolds of ${\cal N}=4$ SYM, by expanding their lagrangian around the local
minimum of the effective potential.  This is a well-posed field theory problem
because the  minimum can be trusted for small coupling.
It would also be very interesting to extend the calculations of \cite{Dymarsky:2005uh, Dymarsky:2005nc} to two loops. At one-loop,
there is no obvious distinction between freely acting and non-freely acting examples.
This distinction may arise at two loops, with the freely acting cases beginning
to show the behavior of Figure \ref{proposal}.

\smallskip

Finally, it would  be nice to find a more detailed  
AdS interpretation for the individual terms appearing in the double-trace beta function. For $\lambda = 0$, when only the term $ v f^2$ is present,
$\beta_f$ can be reproduced by a simple bulk calculation \cite{Witten:2001ua}, using the interpretation \cite{ Witten:2001ua, Berkooz:2002ug}
   of the double-trace deformation
as a mixed  boundary condition for the bulk scalar. There should be
a bulk interpretation for the other terms of $\beta_f$ as well, in particular for the coefficient $a(\lambda)$ which drives the instability.

\label{discussion}

\section*{Acknowledgements}

It is pleasure to thank Igor Klebanov, Andrei Parnachev, Martin Rocek for very useful discussions. We thank Adi Armoni for useful correspondence on orientifold
field theories and for pointing out the omission of the important reference \cite{Armoni:2003va} in the first version of this paper.
The work of L.R. is supported in part by the National Science Foundation Grant No. PHY-0354776 and by the DOE Outstanding Junior Investigator Award.   Any opinions, findings, and conclusions or recommendations
expressed in this material are those of the authors and do not necessarily reflect the views of
the National Science Foundation.

\label{global}




\newpage

\begin{thebibliography}{9}

\bibitem{pertfinite}
A.~Parkes and P.~C. West, {\it {Finiteness in Rigid Supersymmetric Theories}},
  {\em Phys. Lett.} {\bf B138} (1984) 99.
;
P.~C. West, {\it {The Yukawa beta Function in N=1 Rigid Supersymmetric
  Theories}},  {\em Phys. Lett.} {\bf B137} (1984) 371.
;
D.~R.~T. Jones and L.~Mezincescu, {\it {The Chiral Anomaly and a Class of Two
  Loop Finite Supersymmetric Gauge Theories}},  {\em Phys. Lett.} {\bf B138}
  (1984) 293.
;
A.~V. Ermushev, D.~I. Kazakov, and O.~V. Tarasov, {\it {FINITE N=1
  SUPERSYMMETRIC GRAND UNIFIED THEORIES}},  {\em Nucl. Phys.} {\bf B281} (1987)
  72--84.
;
D.~I. Kazakov, {\it {FINITE N=1 SUSY FIELD THEORIES AND DIMENSIONAL
  REGULARIZATION}},  {\em Phys. Lett.} {\bf B179} (1986) 352--354.
;
D.~R.~T. Jones, {\it {COUPLING CONSTANT REPARAMETRIZATION AND FINITE FIELD
  THEORIES}},  {\em Nucl. Phys.} {\bf B277} (1986) 153.
;
R.~Oehme, {\it {REDUCTION AND REPARAMETRIZATION OF QUANTUM FIELD THEORIES}},
  {\em Prog. Theor. Phys. Suppl.} {\bf 86} (1986) 215.
;
C.~Lucchesi, O.~Piguet, and K.~Sibold, {\it {NECESSARY AND SUFFICIENT
  CONDITIONS FOR ALL ORDER VANISHING BETA FUNCTIONS IN SUPERSYMMETRIC
  YANG-MILLS THEORIES}},  {\em Phys. Lett.} {\bf B201} (1988) 241.
;
X.-d. Jiang and X.-j. Zhou, {\it {A CRITERION FOR EXISTENCE OF FINITE TO ALL
  ORDERS N=1 SYM THEORIES}},  {\em Phys. Rev.} {\bf D42} (1990) 2109--2114.

\bibitem{Banks:1981nn}
T.~Banks and A.~Zaks, {\it {On the Phase Structure of Vector-Like Gauge
  Theories with Massless Fermions}},  {\em Nucl. Phys.} {\bf B196} (1982) 189.

\bibitem{Maldacena:1997re}
J.~M. Maldacena, {\it {The large N limit of superconformal field theories and
  supergravity}},  {\em Adv. Theor. Math. Phys.} {\bf 2} (1998) 231--252,
  [\href{http://xxx.lanl.gov/abs/hep-th/9711200}{{\tt hep-th/9711200}}].

\bibitem{Gubser:1998bc}
S.~S. Gubser, I.~R. Klebanov, and A.~M. Polyakov, {\it {Gauge theory
  correlators from non-critical string theory}},  {\em Phys. Lett.} {\bf B428}
  (1998) 105--114, [\href{http://xxx.lanl.gov/abs/hep-th/9802109}{{\tt
  hep-th/9802109}}].

\bibitem{Witten:1998qj}
E.~Witten, {\it {Anti-de Sitter space and holography}},  {\em Adv. Theor. Math.
  Phys.} {\bf 2} (1998) 253--291,
  [\href{http://xxx.lanl.gov/abs/hep-th/9802150}{{\tt hep-th/9802150}}].

\bibitem{Kachru:1998ys}
S.~Kachru and E.~Silverstein, {\it {4d conformal theories and strings on
  orbifolds}},  {\em Phys. Rev. Lett.} {\bf 80} (1998) 4855--4858,
  [\href{http://xxx.lanl.gov/abs/hep-th/9802183}{{\tt hep-th/9802183}}].

\bibitem{Lawrence:1998ja}
A.~E. Lawrence, N.~Nekrasov, and C.~Vafa, {\it {On conformal field theories in
  four dimensions}},  {\em Nucl. Phys.} {\bf B533} (1998) 199--209,
  [\href{http://xxx.lanl.gov/abs/hep-th/9803015}{{\tt hep-th/9803015}}].

\bibitem{Bershadsky:1998mb}
M.~Bershadsky, Z.~Kakushadze, and C.~Vafa, {\it {String expansion as large N
  expansion of gauge theories}},  {\em Nucl. Phys.} {\bf B523} (1998) 59--72,
  [\href{http://xxx.lanl.gov/abs/hep-th/9803076}{{\tt hep-th/9803076}}].

\bibitem{Adams:2001jb}
A.~Adams and E.~Silverstein, {\it {Closed string tachyons, AdS/CFT, and large N
  QCD}},  {\em Phys. Rev.} {\bf D64} (2001) 086001,
  [\href{http://xxx.lanl.gov/abs/hep-th/0103220}{{\tt hep-th/0103220}}].

\bibitem{Dymarsky:2005uh}
A.~Dymarsky, I.~R. Klebanov, and R.~Roiban, {\it {Perturbative search for fixed
  lines in large N gauge theories}},  {\em JHEP} {\bf 08} (2005) 011,
  [\href{http://xxx.lanl.gov/abs/hep-th/0505099}{{\tt hep-th/0505099}}].

\bibitem{Dymarsky:2005nc}
A.~Dymarsky, I.~R. Klebanov, and R.~Roiban, {\it {Perturbative gauge theory and
  closed string tachyons}},  {\em JHEP} {\bf 11} (2005) 038,
  [\href{http://xxx.lanl.gov/abs/hep-th/0509132}{{\tt hep-th/0509132}}].

\bibitem{Bershadsky:1998cb}
M.~Bershadsky and A.~Johansen, {\it {Large N limit of orbifold field
  theories}},  {\em Nucl. Phys.} {\bf B536} (1998) 141--148,
  [\href{http://xxx.lanl.gov/abs/hep-th/9803249}{{\tt hep-th/9803249}}].

\bibitem{Tseytlin:1999ii}
A.~A. Tseytlin and K.~Zarembo, {\it {Effective potential in non-supersymmetric
  SU(N) x SU(N) gauge theory and interactions of type 0 D3-branes}},  {\em
  Phys. Lett.} {\bf B457} (1999) 77--86,
  [\href{http://xxx.lanl.gov/abs/hep-th/9902095}{{\tt hep-th/9902095}}].

\bibitem{Csaki:1999uy}
C.~Csaki, W.~Skiba, and J.~Terning, {\it {Beta functions of orbifold theories
  and the hierarchy problem}},  {\em Phys. Rev.} {\bf D61} (2000) 025019,
  [\href{http://xxx.lanl.gov/abs/hep-th/9906057}{{\tt hep-th/9906057}}].

\bibitem{Breitenlohner:1982bm}
P.~Breitenlohner and D.~Z. Freedman, {\it {Positive Energy in anti-De Sitter
  Backgrounds and Gauged Extended Supergravity}},  {\em Phys. Lett.} {\bf B115}
  (1982) 197.

\bibitem{Lunin:2005jy}
O.~Lunin and J.~M. Maldacena, {\it {Deforming field theories with U(1) x U(1)
  global symmetry and their gravity duals}},  {\em JHEP} {\bf 05} (2005) 033,
  [\href{http://xxx.lanl.gov/abs/hep-th/0502086}{{\tt hep-th/0502086}}].

\bibitem{Frolov:2005dj}
S.~Frolov, {\it {Lax pair for strings in Lunin-Maldacena background}},  {\em
  JHEP} {\bf 05} (2005) 069,
  [\href{http://xxx.lanl.gov/abs/hep-th/0503201}{{\tt hep-th/0503201}}].

\bibitem{Ananth:2007px}
S.~Ananth, S.~Kovacs, and H.~Shimada, {\it {Proof of ultra-violet finiteness
  for a planar non- supersymmetric Yang-Mills theory}},  {\em Nucl. Phys.} {\bf
  B783} (2007) 227--237, [\href{http://xxx.lanl.gov/abs/hep-th/0702020}{{\tt
  hep-th/0702020}}].

\bibitem{paper2}
E.~Pomoni and L.~Rastelli: To appear.

\bibitem{Horowitz:2007pr}
G.~T. Horowitz, J.~Orgera, and J.~Polchinski, {\it {Nonperturbative Instability
  of $AdS_5 x S^5/Z_k$}},  {\em Phys. Rev.} {\bf D77} (2008) 024004,
  [\href{http://xxx.lanl.gov/abs/0709.4262}{{\tt 0709.4262}}].

\bibitem{Copsey:2008fs}
K.~Copsey and R.~B. Mann, {\it {States of Negative Energy and $AdS_5 \times
  S_5/Z_k$}},  {\em JHEP} {\bf 05} (2008) 069,
  [\href{http://xxx.lanl.gov/abs/0803.3801}{{\tt 0803.3801}}].

\bibitem{Aharony:2001pa}
O.~Aharony, M.~Berkooz, and E.~Silverstein, {\it {Multiple-trace operators and
  non-local string theories}},  {\em JHEP} {\bf 08} (2001) 006,
  [\href{http://xxx.lanl.gov/abs/hep-th/0105309}{{\tt hep-th/0105309}}].

\bibitem{Aharony:2001dp}
O.~Aharony, M.~Berkooz, and E.~Silverstein, {\it {Non-local string theories on
  AdS(3) x S**3 and stable non-supersymmetric backgrounds}},  {\em Phys. Rev.}
  {\bf D65} (2002) 106007, [\href{http://xxx.lanl.gov/abs/hep-th/0112178}{{\tt
  hep-th/0112178}}].

\bibitem{Witten:2001ua}
E.~Witten, {\it {Multi-trace operators, boundary conditions, and AdS/CFT
  correspondence}},  \href{http://xxx.lanl.gov/abs/hep-th/0112258}{{\tt
  hep-th/0112258}}.

\bibitem{Berkooz:2002ug}
M.~Berkooz, A.~Sever, and A.~Shomer, {\it {Double-trace deformations, boundary
  conditions and spacetime singularities}},  {\em JHEP} {\bf 05} (2002) 034,
  [\href{http://xxx.lanl.gov/abs/hep-th/0112264}{{\tt hep-th/0112264}}].

\bibitem{Minces:2002wp}
P.~Minces, {\it {Multi-trace operators and the generalized AdS/CFT
  prescription}},  {\em Phys. Rev.} {\bf D68} (2003) 024027,
  [\href{http://xxx.lanl.gov/abs/hep-th/0201172}{{\tt hep-th/0201172}}].

\bibitem{Freedman:1991tk}
D.~Z. Freedman, K.~Johnson, and J.~I. Latorre, {\it {Differential
  regularization and renormalization: A New method of calculation in quantum
  field theory}},  {\em Nucl. Phys.} {\bf B371} (1992) 353--414.

\bibitem{Freedman:1992gr}
D.~Z. Freedman, K.~Johnson, R.~Munoz-Tapia, and X.~Vilasis-Cardona, {\it {A
  Cutoff procedure and counterterms for differential renormalization}},  {\em
  Nucl. Phys.} {\bf B395} (1993) 454--496,
  [\href{http://xxx.lanl.gov/abs/hep-th/9206028}{{\tt hep-th/9206028}}].

\bibitem{Latorre:1993xh}
J.~I. Latorre, C.~Manuel, and X.~Vilasis-Cardona, {\it {Systematic differential
  renormalization to all orders}},  {\em Ann. Phys.} {\bf 231} (1994) 149--173,
  [\href{http://xxx.lanl.gov/abs/hep-th/9303044}{{\tt hep-th/9303044}}].

\bibitem{Coleman:1973jx}
S.~R. Coleman and E.~J. Weinberg, {\it {Radiative Corrections as the Origin of
  Spontaneous Symmetry Breaking}},  {\em Phys. Rev.} {\bf D7} (1973)
  1888--1910.

\bibitem{Beisert:2005he}
N.~Beisert and R.~Roiban, {\it {The Bethe ansatz for Z(S) orbifolds of N = 4
  super Yang- Mills theory}},  {\em JHEP} {\bf 11} (2005) 037,
  [\href{http://xxx.lanl.gov/abs/hep-th/0510209}{{\tt hep-th/0510209}}].

\bibitem{princeton}
A.~Dymarsky, S.~Franco, I.~R. Klebanov, and R.~Roiban: Work in progress.

\bibitem{Armoni:2003va}
A.~Armoni, E.~Lopez, and A.~M. Uranga, {\it {Closed strings tachyons and
  non-commutative instabilities}},  {\em JHEP} {\bf 02} (2003) 020,
  [\href{http://xxx.lanl.gov/abs/hep-th/0301099}{{\tt hep-th/0301099}}].

\bibitem{Frampton:1999yb}
P.~H. Frampton and C.~Vafa, {\it {Conformal approach to particle
  phenomenology}},  \href{http://xxx.lanl.gov/abs/hep-th/9903226}{{\tt
  hep-th/9903226}}.

\bibitem{Frampton:2007fr}
P.~H. Frampton and T.~W. Kephart, {\it {Quiver Gauge Theory and Conformality at
  the Large Hadron Collider}},  {\em Phys. Rept.} {\bf 454} (2008) 203--269,
  [\href{http://xxx.lanl.gov/abs/0706.4259}{{\tt 0706.4259}}].

\bibitem{Kakushadze:2007qx}
Z.~Kakushadze, {\it {Tachyon-Free Non-Supersymmetric Strings on Orbifolds}},
  {\em Int. J. Mod. Phys.} {\bf A23} (2008) 4371--4386,
  [\href{http://xxx.lanl.gov/abs/0711.4108}{{\tt 0711.4108}}].

\bibitem{Armoni:2007jt}
A.~Armoni, {\it {Non-Perturbative Planar Equivalence and the Absence of Closed
  String Tachyons}},  {\em JHEP} {\bf 04} (2007) 046,
  [\href{http://xxx.lanl.gov/abs/hep-th/0703229}{{\tt hep-th/0703229}}].

\end{thebibliography}
\end{document}